\documentclass{article}
\date{}

\usepackage[margin=1.in]{geometry}

\usepackage{natbib}

\usepackage{graphicx}
\usepackage{amsmath,amsthm,bm,mathrsfs,amsfonts}
\usepackage{xcolor}

\usepackage{tikz}
\usetikzlibrary{calc, positioning, shapes, through, intersections}
\usepackage[normalem]{ulem} %to strike the words
\usetikzlibrary{positioning,fit}
\usetikzlibrary{decorations.text}
\usetikzlibrary{arrows}
\usetikzlibrary{shapes}

 \newtheoremstyle{theorem}{6pt}{6pt}{\rm}{}{\sffamily}{ }{ }{}
 \theoremstyle{theorem}
\newtheorem{theorem}{\sc Theorem}[section]

 \newtheoremstyle{algorithm}{6pt}{6pt}{\rm}{}{\sffamily}{ }{ }{}
 \theoremstyle{algorithm}

 \newtheoremstyle{lemma}{6pt}{6pt}{\rm}{}{\sffamily}{ }{ }{}
 \theoremstyle{lemma}

\newtheoremstyle{case}{6pt}{6pt}{\rm}{}{\sffamily}{. }{ }{}
 \theoremstyle{case}

 \newtheoremstyle{statement}{6pt}{6pt}{\rm}{}{\sffamily}{ }{ }{}
\theoremstyle{statement}

 \newtheoremstyle{corollary}{6pt}{6pt}{\rm}{}{\sffamily}{ }{ }{}
 \theoremstyle{corollary}

  \newtheoremstyle{definition}{6pt}{6pt}{\rm}{}{\sffamily}{ }{ }{}
 \theoremstyle{definition}

\newtheoremstyle{example}{6pt}{6pt}{\rm}{}{\sffamily}{ }{ }{}
\theoremstyle{example}
\newtheorem{example}[theorem]{\sc Example}

\newtheoremstyle{remark}{6pt}{6pt}{\rm}{}{\sffamily}{ }{ }{}
\theoremstyle{remark}

\newtheoremstyle{approximation}{6pt}{6pt}{\rm}{}{\sffamily}{ }{ }{}
\theoremstyle{approximation}

\newtheoremstyle{scheme}{6pt}{6pt}{\rm}{}{\sffamily}{ }{ }{}
\theoremstyle{scheme}

\newtheoremstyle{Algorithm}{6pt}{6pt}{\rm}{}{\sffamily}{ }{ }{}
\theoremstyle{Algorithm}

\newtheoremstyle{Assumption}{6pt}{6pt}{\rm}{}{\sffamily}{ }{ }{}
\theoremstyle{Assumption}

\newtheoremstyle{proposition}{6pt}{6pt}{\rm}{}{\sffamily}{ }{ }{}
\theoremstyle{proposition}

\newtheoremstyle{hypo}{6pt}{6pt}{\rm}{}{\sffamily}{ }{ }{}
 \theoremstyle{hypo}

  \newtheoremstyle{Step}{6pt}{6pt}{\rm}{}{}{ }{ }{}
 \theoremstyle{Step}

\numberwithin{equation}{section}

\DeclareMathOperator{\EX}{\mathbb{E}}% expected value

\newcommand{\rev}[1]{\textcolor{black}{#1}}

\begin{document}

\title{Optimal sports betting strategies in practice:\\ an experimental review \vspace{-1cm}}
% \author{ {\sc Matej Uhrin}\\[2pt]
% Czech Technical University, Prague\\[6pt]}
% \pagestyle{headings}

\maketitle
% \author{}

\author{ \centering {\sc Matej Uhr\'{i}n, Gustav \v{S}ourek, Ond\v{r}ej Hub\'{a}\v{c}ek, Filip \v{Z}elezn\'{y}}\\[2pt]
\centering \textit{Department of Computer Science}\\[2pt]
\centering \textit{Czech Technical University in Prague}\\[16pt]
% Czech Technical University in Prague
 }

% Institutes for affiliations are also joined by \and,
% \institute{
%   Czech Technical University, Prague
%  }

%  \authorrunning{} has to be set for the shorter version of the authors' names;
% otherwise a warning will be rendered in the running heads. When processed by
% EasyChair, this command is mandatory: a document without \authorrunning
% will be rejected by EasyChair

%%%%%%%%%%%%%%%%%abstract style
%Two grouping braces are necessary in abstract environment
%first argument contains abstract text; second argument contains keywords
%text

% \begin{abstract}
% {We investigate the most prominent streams of approaches to the problem of sports betting investment based on the Modern portfolio theory and the Kelly criterion. We define the problem settings, the formal strategies, and review their modifications for additional risk control stemming from \rev{their} unrealistic mathematical assumptions that are not met in betting practice. We test the resulting methods following a unified evaluation protocol in 3 different sport\rev{s} domains of horse racing, basketball and football. The results are generally in \rev{favour} of the formal approaches while suggesting for the individual benefits of the additional risk control practices together with their trade-offs.}
% {sports betting, betting strategies, risk management, bankroll management}
% \end{abstract}

\begin{abstract}
{We investigate the most popular approaches to the problem of sports betting investment based on modern portfolio theory and the Kelly criterion. We define the problem setting, the formal investment strategies, and review their common modifications used in practice. The underlying purpose of the reviewed modifications is to mitigate the additional risk stemming from the unrealistic mathematical assumptions of the formal strategies. We test the resulting methods using a unified evaluation protocol for three sports: horse racing, basketball and soccer. The results show the practical necessity of the additional risk-control methods and demonstrate their individual benefits. Particularly, an adaptive variant of the popular ``fractional Kelly'' method is a very suitable choice across a wide range of settings.}
{sports betting, betting strategies, risk management, bankroll management}
\end{abstract}
%%%%%%%%%%%%%%%%%%%%%%%%%%%%%%%

%%%%%%%%%%%%%%section A%%%%%%%%%
\section{Introduction}
\label{sec:intro}

Sports betting systems generally consist of two essential components \rev{--} (i) predictive models, generating probabilistic estimates for the given match outcomes, and (ii) bankroll management strateg\rev{ies}, optimizing the expected progression of wealth in time. In this work, we focus solely on the latter.

While much of the available research on betting systems is \rev{centred} around the predictive \rev{modelling} part, often completely neglecting the need for betting portfolio optimization, we show that, given a predictive model, the betting strategy has a major influence on the final measures of profit. Consequently, a worse model with a better strategy can easily outperform a better model with a worse strategy.

Lacking a deeper understanding of the investment part of the problem, practitioners often resort to trivial practices such as various forms of flat betting. We show that these are inferior to the formal strategies, not just theoretically but also from \rev{a} practical perspective. There are two basic streams of research in the formal approaches, stemming from information theory and economics, respectively. The first, and the most widespread, is the Kelly criterion\rev{~\citep{kelly1956new}}, also known as the geometric mean policy, maximizing the expected long-term growth of wealth. The second is the approach of Markowitz's Modern portfolio theory\rev{~\citep{markowitz1952portfolio}}, balancing the criteria of expected profit and \rev{profit} variance as a measure of risk.

While mathematically sound, the formal strategies are based on unrealistic assumptions. The most limiting assumption in their application to sports betting is the knowledge of true probabilities of individual match outcomes. Other complications of the problem include \rev{the} multiplicity of outcomes and parallel matches. We investigate the existing modifications of the formal strategies proposed to address the problems occurring in practice and evaluate them experimentally in 3 different sport\rev{s} domains - horse racing, basketball, and football.

The paper is structured as follows. In Section~\ref{sec:definitions} we define the concept of a betting strategy and the dimensions of the underlying optimization problem. In Section~\ref{sec:related} we review the related work touching different facets of risk and bankroll management in betting. In Section~\ref{sec:strategies} we formally introduce the two core strategies of Kelly and Markowitz. The modifications of the core strategies proposed to manage the extra risk occurring in practice are then introduced in Section~\ref{sec:risk}. Finally, we experimentally evaluate the strategies in practical scenarios in Section~\ref{sec:experiments} and conclude the paper in Section~\ref{sec:conclusion}.

\section{Problem Definition}
\label{sec:definitions}

In its core, sports betting is a simple stochastic game where the player $p$ repeatedly allocates a distribution of \textit{fractions} ($f_i \in [0,1],~\sum_{i}f_i \leq 1$) of her current bankroll $W \in \mathbb{R}$ at time $t \in \mathbb{N}$ over possible stochastic results $r_i \in \mathrm{R}$ of a match, coming from a distribution $P_r(r_i)$ over the domain $\mathrm{R}$ of the random variable $R$, describing all the possible outcomes of the given match at time step $t$. Each of the possible match outcomes $r_i$ is then associated with \rev{so-called} \textit{odds} ($o_i \in \mathbb{R}_{\geq 1}$) by the bookmaker $b: r_i \mapsto o_i$. Should a particular outcome $i$ be realized \rev{(}${R}=r_i$\rev{)}, a payoff $o_i \cdot f_i \cdot W$ from the associated odds and fraction is to be received by the player $p$. In the opposite case, the player loses the allocated portion $f_i \cdot W$ of her bankroll to the bookmaker $b$.
Each of the particular \rev{betting} outcomes $r_i$ is \rev{thus} binary\footnote{\rev{Note this concerns an individual $r_i$ and not $|\mathrm{R}|$, i.e. a match can have many possible outcomes $r_i$, but each $r_i$ has a binary realization, resulting exclusively in either win or loss of the bet associated with it.}} in nature, and the potential net profit $w_i$ from allocation on the $i$-th outcome is thus
\begin{equation}
    w_i =
\left\{
	\begin{array}{lll}
		o_i \cdot f_i \cdot W - f_i \cdot W ~~& \mbox{with prob. $P_r(r_i)$}  &\mbox{(if $\mathrm{R}=r_i$ is realized)} \\
        - f_i \cdot W   ~~& \mbox{with prob. $1-P_r(r_i)$}  &\mbox{(if $\mathrm{R} \neq r_i$)}
	\end{array}
\right.
\end{equation}
giving an expectation
\begin{equation}
    \EX_{P_r}[w_i] = P_r(r_i) \cdot (o_i f_i  W - f_i  W) + (1-P_r(r_i)) \cdot (- f_i  W)
\end{equation}

Clearly, the profits of the bettor and bookmaker are directly opposite and, assuming a closed system of bettors and bookmakers, this is a zero-sum game. The goal of both the player $p$ and the bookmaker $b$ is to maximize their long-term profits $W_{t \to \infty}$ as measured by their respective utilities (Section~\ref{sec:strategies}). Given the stochastic nature of the game, the natural desideratum of the player is to allocate the fractions $\bm{f} = f_1, \dots, f_n$ so as to target a high total expect\rev{ation of profit} $\mathrm{W}$

\begin{equation}
    \EX_{P_r}[\mathrm{W}] = \EX_{P_r} \bigg[\sum_i w_i \bigg] = \sum_i \EX_{P_r} [w_i]
\end{equation}

Note that, in this work, we assume the two players to take on the asymmetric roles of market maker $b$ and market taker $p$, where the bookmaker $b$ always starts by laying out the odds $\bm{o} = [o_1, \dots, o_n]$ for the possible match results $\bm{r} = [r_1, \dots, r_n]$ first, consequently to which the player $p$ reacts with his best policy for allocation $p : r_i \mapsto f_i$ of her current wealth $W_t$. In contrast to e.g. the, \rev{less common}, betting exchange setting, in this work we assume solely the strategies for the role of the market taker $p$, which is the most common setup for bettors in practice.

\subsection{Betting Strategy}
\label{sec:def:strategy}

A player's betting strategy for a game with $n$ outcomes is a \rev{function $g$} mapping a set of probabilistic estimates $\hat{\bm{p}} = \hat{p_i}, \dots,\hat{p_n}$ and bookmaker's odds $\bm{o} = o_1, \dots, o_n$ onto a set of fractions $\bm{f} = f_1, \dots, f_n$ of the current wealth $W_t$ to be waged \rev{over} the game outcomes $\bm{r} = r_1, \dots, r_n$
\rev{
\begin{align}
 g &: (\hat{\bm{p}}, \bm{o}) \mapsto \bm{f} 
\end{align}
}
Typically, the estimated distribution vector $\hat{\bm{p}}$ comes from a probabilistic model $P_p$ of the player and is similar to, yet \rev{most likely} different from, the \rev{(unknown)} true probability distribution $P_p = \hat{P_r},~P_p \neq P_r$ \rev{(Section \ref{sec:def:estimates})}.
The vector of the waged fractions $\bm{f}$ is then often referred to as the \textit{portfolio} over individual ``assets'' $i$ (Section~\ref{sec:MPT})
\begin{equation}
\bm{f} =
\begin{bmatrix}
   f_1, \dots, f_n
\end{bmatrix}
\end{equation}
where $f_i$ indicates the portion of wealth $W_t$ allocated to $i$-th outcome.

\subsection{Fixed Odds}
\label{sec:def:odds}

We further assume a \rev{so-called} fixed-odds betting setup which, as opposed to e.g. parimutuel setting~\citep{hausch2008efficiency}, always offers known odds distribution $\bm{o}$ in advance of the game for the player's strategy \rev{$g$} to calculate with.
In its most basic form, we can demonstrate the given setting on a simple \rev{coin-tossing} game as follows.

\begin{example}
\label{ex:coin1}
Assume a fair \rev{coin-tossing} game with two, equally probable, outcomes $\mathrm{R} =\{Heads, Tails\}$
\begin{equation}
    \underset{r_i \in \mathrm{R}}{P_r(r_i)} =
\left\{
	\begin{array}{ll}
		0.5  & \mbox{for the coin falling } r_1 = \textit{Heads} \\
		0.5  & \mbox{for the coin falling } r_2 = \textit{Tails}
	\end{array}
\right.
\end{equation}
The odds by the bookmaker $b$ could then be set up e.g. as follows
\begin{equation}
    \underset{r_i \in \mathrm{R}}{b(r_i)} =
\left\{
	\begin{array}{ll}
		o_1 = 1.9  & \mbox{for the coin falling } r_1 = \textit{Heads} \\
		o_2 = 1.9  & \mbox{for the coin falling } r_2 = \textit{Tails}
	\end{array}
\right.
\end{equation}
Let the bettor allocate a fixed wager, such as \$1, on the $r_1=Heads$.
% \footnote{Note that she could as well choose \textit{Tails} or both or neither.}. 
She then receives an extra $w_1 = (1.9 - 1) * 1$ profit if the associated outcome $r_1=Heads$ is realized, or losses the placed wager \$1 otherwise.
It is easy to see that this particular game is generally disadvantageous for the bettor, and there exist no strategy for her to make long-term profits, since the expected profit for each outcome of the game is simply negative:
\begin{equation}
\EX[w_1] = \EX[w_2] = 0.5 \cdot 1.9 \cdot 1 + 0.5 \cdot (-1) = -0.05
\end{equation}
This \rev{follows directly from} the fact that the odds are \textit{unbiased} and \textit{subfair}. This means that \rev{the distribution of their inverse values $P_b : r_i \mapsto \frac{1}{o_i}$ is} proportional to the true probability distribution over the game outcomes, but \rev{it does} not form a \textit{probability} distribution as \rev{the values} do not sum up to $1$:

\begin{equation}
    \sum_i{P_b(r_i)} = \frac{1}{o_1} + \frac{1}{o_2} \approx 1.05
\end{equation}
\end{example}
% \begin{equation}
%     odds :
% \left\{
% 	\begin{array}{ll}
% 		fair :&\sum_{i=1}^k\frac{1}{o_i} = 1.0 \\
% 		subfair :&\sum_{i=1}^k\frac{1}{o_i} > 1.0 \\
% 		superfair :&\sum_{i=1}^k\frac{1}{o_i} < 1.0
% 	\end{array}
% \right.
% \end{equation}
Out of the three settings~\citep{cover2012elements}: \textit{fair, subfair, superfair}, 
the \textit{subfair} odds are typically the only setting for a bookmaker to be able to generate profits. We will further limit ourselves to this setting as it is the only valid setup working in practice.
The value of 
\rev{
\begin{equation}
margin = \frac{\sum_{j=1}^K\frac{1}{o_j} -1 }{\sum_{j=1}^K\frac{1}{o_j}}
\end{equation}}
is then called the bookmaker's margin\footnote{Often wrongly calculated as simply the remainder \rev{over $1$ as  $\sum_{j=1}^K\frac{1}{o_j} -1$}} (also known as ``vigorish'', ``house edge'', ``cut'' etc.), and represents the negative expected value of the game given the probabilities $P_b$ implied from the odds are unbiased estimates of the true outcome probabilities $P_r$. Note that this is a typical game setting operated in the gambling industry, such as in various casino games, where there is no space for long-term profitable strategies. However, we note that the situation in sports betting is principally different.

\subsection{Biased Estimates}
\label{sec:def:estimates}

In Example~\ref{ex:coin1} with a fair coin, both the bookmaker and bettor knew the true outcome probability distribution (i.e. $P_r(r_1=H)=0.5 ;\ P_r(r_2=T)=0.5$). This setting is very elegant from mathematical perspective, since one can calculate exact expected values of profits and consequently derive optimal betting strategies (Section~\ref{sec:strategies}).
Such mathematically optimal strategies can be theoretically applied in artificial environments with handcrafted generators of randomness (e.g. the casinos). However, in the context of sports betting, and other practical settings such as stock market investing, this is generally impossible.
In this experimental review, we thus focus on the scenarios, where the probability estimates of both the bookmaker $P_b$ and the player $P_p$ are biased w.r.t. the real outcome probability distribution $P_r$.
Let us consider an extension of the \rev{coin-tossing} game from Example~\ref{ex:coin1} to demonstrate properties of such \rev{a} setting.
\begin{example}
Consider a \textit{biased} \rev{coin-tossing} game where the coin bias is \textit{unknown} to both the bookmaker and the player. Let us \rev{set-up} the bias such that
\begin{equation}
\underset{r_i \in {\mathrm{R}}}{P_r(r_i)} =
\left\{
	\begin{array}{ll}
		0.6  & \mbox{for } r_1 = \textit{H} \\
		0.4  & \mbox{for } r_2 = \textit{T}
	\end{array}
\right.
\end{equation}
Let us further assume that the player $p$ has a probabilistic model of the game, producing biased estimates $P_p = \hat{P_r}$ as
\begin{equation}
\underset{r_i \in {\mathrm{R}}}{P_p(r_i)} =
\left\{
	\begin{array}{ll}
		0.55  & \mbox{for } r_1 = \textit{H} \\
		0.45  & \mbox{for } r_2 = \textit{T}
	\end{array}
\right.
\end{equation}
Finally, assume the bookmaker is also biased with his estimates $P_b = \hat{P_r}, P_b \neq P_p$, according to which he sets up the odds distribution $\bm{o}$, lowered by a margin\footnote{In practice, the distribution of margin would not be simply uniform as in the example, but the bookmaker typically applies more sophisticated distortion of the odds to secure even higher statistical advantage.} $m=0.05$
\begin{equation}
\underset{r_i \in {\mathrm{R}}}{P_b(r_i)} =
\left\{
	\begin{array}{ll}
		0.65  & \mbox{for } r_1 = \textit{H} \\
		0.35  & \mbox{for } r_2 = \textit{T}
	\end{array}
\right.
\underset{r_i \in {\mathrm{R}}}{b(r_i)} =
\left\{
	\begin{array}{ll}
		\frac{1}{0.65} \cdot (1-{0.05}) \approx 1.46  & \mbox{for } r_1 = \textit{H} \\
		\frac{1}{0.35} \cdot (1-{0.05})  \approx 2.71 & \mbox{for } r_2 = \textit{T}
	\end{array}
\right.
\end{equation}
Note that while the odds are still subfair, the bookmaker's bias w.r.t. $P_r$ now creates space for exploitation, since the true expected values are no longer purely negative.
\begin{equation}
	\begin{array}{llll}
		\EX_{P_r}[w_1] &=& P_r(r_1) \cdot b(r_1) -1 \approx -0.124 & \text{ for~~   } \mathrm{R}=r_1=H\\
		\EX_{P_r}[w_2] &=& P_r(r_2) \cdot b(r_2) -1 \approx 0.084 & \text{ for~~  } \mathrm{R}=r_2=T
	\end{array}
\end{equation}
i.e. the punter could make long-term profits if betting appropriate amounts on the $r_2=T$ outcome. However, not knowing the true probabilities $P_r$, the player's calculation of expected values will now be biased, too
\begin{equation}
	\begin{array}{lll}
		\EX_{P_p}[w_1] &=& P_p(r_1) \cdot b(r_1) -1 \approx -0.197\\
		\EX_{P_p}[w_2] &=& P_p(r_2) \cdot b(r_2) -1 \approx 0.22
	\end{array}
\end{equation}
nevertheless, despite the expected values calculated by the punter w.r.t. her $P_p$ estimate \rev{being} wrong, in this particular setting, she correctly identified the positive expected value in the $r_2=T$ outcome and could theoretically make a profit with an appropriate strategy modification (Section~\ref{sec:risk}).
\end{example}

\begin{figure}[t]
\label{fig:triangle}
\centering
\includegraphics{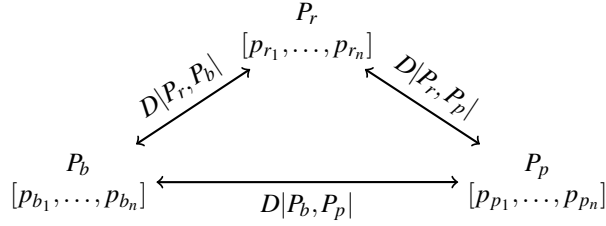}
\caption{A typical sports betting setting for a game with $n$ outcomes, displaying bookmaker's probabilistic estimates $P_b$ and player's estimates $P_p$, both distanced from the true distribution $P_r$ and from each other.}
\end{figure}

Generally, $P_p = \hat{P_r}$ and $P_b = \hat{P_r}^{'}$ are going to be somewhat biased w.r.t. $P_r$ as well as w.r.t. each other \rev{(i.e. $P_p \neq P_b$,} as long as \rev{the} player does not simply copy from the bookmaker). The individual biases can be captured by statistical measures, such as the Kullback\-Leibler, or better yet Jensen\-Shannon, divergences~\citep{cover2012elements}, and the probabilistic setting of each game for a particular match can then be understood as a triplet of probability distributions over the outcomes, as depicted in Figure~\ref{fig:triangle}.

\subsection{Multiplicity of Outcomes}
\label{sec:def:outcomes}

So far we have assumed a binary \rev{coin-tossing} game of two possible outcomes. Let us now generalize into an $n$ outcome game, such as throwing a die. This represents most real situations in sports betting, such as the $\mathrm{R} = \{Win,Draw,Loss\}$ outcomes in soccer, or betting on the winner of a horse race with $n$ horses (Section~\ref{sec:datasets}). Moreover, one can potentially assume that the individual game outcomes are no longer exclusive, such as betting on the first $j$ horses, or ``over'' $j$ goals in soccer for multiple different values of $j$.

To make the game representation more compact in such situations, a generic matrix~$\bm{O}$ representation has been proposed~\citep{busseti2016risk}, where the columns of $\bm{O}$ represent the possible outcome assets, and rows represent the possible game results, i.e. joint realizations of all the outcomes. Each individual element in $\bm{O}$ then represents particular odds for each outcome realization.

Additionally, we include an artificial risk-free ``cash'' asset $\bm{c}$, which allows the player to put money aside safely. This also allows to model situations where leaving money aside can cost \rev{a} small fraction of wealth in every turn (caused \rev{e.g.} by inflation), or possibility to increase the wealth by some interest rate (e.g. in a savings account).

The betting strategy \rev{$g$} (Section~\ref{sec:def:strategy}) can now thus always allocate the full amount of current wealth $W$ among $n$ available outcome assets, $n - 1$ of which are risky, stochastic assets, and 1 being the added risk-free cash asset as
\begin{equation}
    g : (\bm{p}^k, \bm{O}_k^n) \mapsto \bm{f}^n \text{~~~where~~~} \sum_i{f_i}=1
\end{equation}
where $k$ is the number of possible worlds, i.e. there are $k$ possible joint outcome realizations, in our probabilistic game.
Odds for each outcome asset in each of the $k$ world realizations with the respective probabilities $\bm{p} = p_1, p_2, ..., p_k$ can thus be fully specified in the columns $\bm{o_i}$ as
\begin{align}
\bm{O} =
\begin{bmatrix}
   \bm{o_1} & \bm{o_2} & ... & \bm{o_{n-1}} & \bm{c}
\end{bmatrix}
~,~\text{where}~~
\bm{o_i} =
    \begin{bmatrix}
       o_{i, 1} \\ 
       o_{i, 2} \\ 
       ... \\ 
       o_{i, n}
    \end{bmatrix}
~,~
\bm{c} =
    \begin{bmatrix}
       1 \\ 
       1 \\ 
       ... \\ 
       1
    \end{bmatrix} 
\end{align}

\begin{example}
Consider a football game, where we assume $3$ outcomes as $\mathrm{R} = \{W, D, L\}$, forming the $3$ asset vectors $\bm{o_w}, \bm{o_d}, \bm{o_l}$, where the bookmaker sets the odds to $o_w, o_d, o_l$, respectively. The odds matrix $\bm{O}$, including the constant cash asset $\bm{c}$, then looks as follows.

\begin{equation}
    \bm{O} =
\begin{bmatrix}
   \bm{o_w} & \bm{o_d} & \bm{o_l} & \bm{c}
\end{bmatrix}
~~\text{where~}~~
\bm{o_w} =
    \begin{bmatrix}
       o_w \\ 
       0 \\ 
       0
    \end{bmatrix}
,~
\bm{o_d} =
    \begin{bmatrix}
       0 \\ 
       o_d \\ 
       0
    \end{bmatrix}
,~
\bm{o_l} =
    \begin{bmatrix}
       0 \\ 
       0 \\ 
       o_l
    \end{bmatrix}
,~
\bm{c} =
    \begin{bmatrix}
       1 \\ 
       1 \\ 
       1
    \end{bmatrix} 
\end{equation}
\end{example}

To simplify notation in further sections, we will also define a modified odds matrix $\bm{\rho}$ corresponding to excess odds, i.e. removing the return amount of the placed wager itself, resulting \rev{in} net profit $\mathrm{W}$ (Section~\ref{sec:definitions}), as
\begin{equation}
    \bm{\rho} = \bm{O} - \bm{1}
\end{equation}
% \begin{equation}
% \bm{\rho} =
% \left\{
% 	\begin{array}{ll}
% 		\bm{O}(i,j) - 1 & \mbox{ where }  \bm{O}(i,j) \geq 1\\
% 		0   & \mbox{ otherwise ~~(i.e. where $\bm{O}(i,j) = 0$)}
% 	\end{array}
% \right.
% \end{equation}

Note that in the example scenario the outcomes were exclusive, and the ``one-hot'' risky asset vectors reflect their exclusive nature, which considerably simplifies the computation of optimal strategies (Section~\ref{sec:strategies}).
In this review, we generally assume individual matches with exclusive outcomes\footnote{Note that the exclusiveness of outcomes does not hold in the further scenarios with parallel games.} but varying outcome multiplicities (Section~\ref{sec:datasets}) to experimentally assess the properties of the strategies w.r.t. this dimension of the problem.

% The player will still search for individual outcomes with positive expected value, and the bookmaker will try to prevent that from happening by approaching the true $P_r$. For an $n$-ary game with fair odds and biased estimates, there will always be at least one outcome with positive expected value w.r.t. each of the distributions, and at least one with a negative expected value. The bookmaker again counters that by additionally applying his margin $m$ to the odds.

\subsubsection{Parallel Games}
\label{sec:def:parallel}

To further complicate the game, approaching the real betting setting even more closely, we can consider multiple dice being thrown in parallel, each associated with a particular set of outcomes and odds. Naturally, this reflects the reality of multiple games being open for betting at the same time. In popular sports, such as soccer, it is not uncommon to have dozens of games available on the market simultaneously.

While we can surely consider each of the games separately, such a simplification can lead to sub-optimal results. Although calculating with the true parallel nature of the opportunities can be computationally demanding for some of the strategies (Section~\ref{sec:quadraticapprox}), it should allow to alleviate the risk by diversifying over a wider portfolio at each time step of the wealth progression.
In this review, we consider both the sequential and parallel scenarios to emulate realistic scenarios and evaluate the respective advantages (Section~\ref{sec:experiments}).

\subsection{Betting Dynamics}
\label{sec:def:dynamics}

The betting dynamic represents the investment \rev{behaviour} of the bettor w.r.t. her bankroll $W$ in time $t$, which has a major impact on the progression of wealth. There are two basic cases of bankroll management to be considered \rev{--} (i) additive and (ii) multiplicative~\citep{peters2016evaluating, peters2011optimal}.

\subsubsection{Additive dynamic}
Additive dynamic corresponds to a simple fixed unit-based investment, where the bettor's wagers are decoupled from her current bankroll $W_t$. To illustrate the setting, we can imagine that the bettor receives a fixed unit (e.g. \$1) amount of money from an external source at regular time intervals $\delta t$ (such as a salary), which she repeatedly invests into the stochastic game of betting, and accumulates (additively) the prospective returns $w_t \cdot 1$ from the unit investment in the, separately held, bankroll $W_t$.
Her wealth progression in time $t$ can hence be seen as
\begin{equation}
    W_t = w_t \cdot 1 + W_{t - \delta t}
\end{equation}

\subsubsection{Multiplicative dynamic}
\label{sec:multiplicative}
In the multiplicative scenario, the bettor continuously \textit{reinvests} the current wealth $W_t$ accumulated from the previous betting investments, without any external source of profit. Hence her progression of wealth in time $t$ can be seen as
\begin{equation}
    W_t = w_t \cdot W_{t - \delta t}
\end{equation}
The multiplicative dynamics plays an important role in the Kelly criterion (Section~\ref{sec:kelly}), where the mathematical optimality of the strategy is derived exactly from \rev{a} repeated play of the same game in the multiplicative setting.

As the comparison of the two approaches appears problematic, due to the external source of profit in the additive scenario, we will further consider only the multiplicative reinvestment setting, which is also more realistic and sound for \rev{an} independent evaluation.

\section{Related works}
\label{sec:related}

The two most notable approaches to allocation of wealth across presented stochastic assets, i.e. match outcomes in sport\rev{s} betting, were introduced by (i)~\cite{markowitz1952portfolio}, with his revolutionary concept of balancing return and risk of a portfolio, and by (ii)~\cite{kelly1956new}, with a criterion to maximize the long-term growth in a scenario where the same game is being played repeatedly.

Following the Kelly criterion, the process of betting is closely connected to information theory~\citep{kelly1956new}. \rev{\cite{bell1988game}, discuss a game-theoretical optimality of Kelly portfolios and a generalization of the Kelly strategy to maximize the proportion of wealth relative to the total wealth among population is discussed in~\citep{lo2018growth}.} Additional mathematical properties were also explored in~\citep{latane2011criteria} and~\citep{breiman1961optimal, thorp2008kelly}. From the economical perspective, Kelly's approach is often explained through the use of a logarithmic utility function, which was famously first introduced by Daniel Bernoulli in~\citep{bernoulli2011exposition}, where he pointed out that people do not make their decisions according to the absolute payoff, but w.r.t. the logarithm thereof. \rev{In~\citep{luenberger2011preference} the authors suggest that assuming long-term goals, the logarithmic utility function is the only sensible choice for a utility function.} While not necessarily incorrect, the phenomenological explanation of the choice of logarithmic utility function seem\rev{s} somewhat arbitrary, however.

In \citep{peters2011time} a different view on the Kelly criterion was proposed, where the author criticized the established evaluation of betting using the expected value of a portfolio, as it is based on the unrealistic idea of ``simultaneous'' evaluation of the, often exclusive, outcomes. Instead of measuring \rev{the} mean of a statistical ensemble of possible outcomes, the author proposed to focus on what happens to a single player as the same game is repeated in time, following the notion of ergodicity in dynamic systems~\citep{peters2019ergodicity}. The logarithmic transformation then emerges as the correct ergodic transformation of dynamics of the game in the classical reinvestment setting~\citep{peters2016evaluating}, providing a well-founded explanation for the observed phenomenon.
Given the mathematically elegant yet somewhat unrealistic setting, the Kelly strategy has also been often criticised in many works~\citep{samuelson1971fallacy, samuelson2011we, maclean2010good, samuelson1975lifetime}.

\subsection{Extensions of the formal strategies}
\label{sec:related:extensions}

The strategies of Markowitz and Kelly have been re-explored by researchers in a number of different application scenarios and many useful modifications have been proposed since. Generally, the Markowitz's approach has traditionally dominated the world of quantitative finance, while the Kelly's approach has been more prominent in the sports betting industry. In~\citep{smoczynski2010explicit}, a closed form solution for the use of the Kelly strategy when betting on horse racing was explored. Another practical extension for betting on multiple simultaneous games was discussed in a number of works~\citep{whitrow2007algorithms, grant2008optimal, buchen2012comparison}, where \rev{various} approximations for large bet aggregations were proposed.

\rev{
Modification of the Kelly strategy for betting exchanges is discussed in~\citep{noon2013kelly}, where adjustments for both back and lay bets are presented. Additionally, the effect of commission and maximum bet constraint on resulting growth rate is discussed. The Kelly problem is examined for spread betting in~\citep{chapman2007kelly} and in \citep{haigh2000kelly}, where several counterintuitive effects are discussed when using the Kelly strategy for spread betting. Markowitz's modern portfolio theory for soccer spread betting is then discussed in~\citep{fitt2009markowitz}
}

Another important stream of research are works investigating extensions of the Kelly strategy towards the realistic setting of parameter uncertainty, such as~\citep{baker2013optimal}.  A practical method to address the problem are \rev{so-called} fractional Kelly strategies, the properties of which have been investigated in great detail in the works of~\citep{maclean2011medium} and \citep{maclean1992growth}.\rev{\cite{peterson2017kelly}, presents a decoupled Kelly strategy combined with an additional risk measure. \cite{kan2007optimal},~introduced an optimal portfolio choice under parameter uncertainty for the modern portfolio theory (MPT).}
Interesting modifications with similar aims are Bayesian extensions of the Kelly strategy proposed in \citep{browne1996portfolio, balka2017kelly, chu2018modified}. Similarly, approaches based on probabilistic risk constraints for limiting the probability of a ``drawdown'' were discussed in \citep{busseti2016risk} and \citep{mulvey2011dynamic}. Finally, limiting the \rev{worst-case} probabilistic scenario using the framework of distributionally robust optimization was explored in \citep{sun2018distributional} and in \citep{blanchet2018distributionally} for the Markowitz's strategy, respectively.

% \todo{
% \textbf{Using statistics to detect match fixing in sport.} Suspicions arise where model odds and market odds diverge. We provide real examples of monitoring for football and tennis matches and describe how suspicious matches are investigated by analysts before a final assessment of how likely it was that a fix took place is made.~\citep{forrest2019using}
% }

\subsection{Predictive modelling}
\label{sec:related:model}
\rev{
Since we consider predictive sports modelling a separate problem, we only briefly review some papers on the topic, with an extra focus on models related to those used for experiments in this paper.
}

\rev{
A traditional stream of research in predictive sports analytics are score-based models based on various explicit statistical assumptions. A football prediction model introduced by~\cite{maher1982}, builds a statistical model on the assumption that in a football match the goals are Poisson-distributed and those of the home team are independent of those of the away team. The author also introduced the notion of teams' attacking and defensive strengths and how to use them for forecasting of the match results. In~\citep{dixon1997}, the Maher's model is further extended and it is shown to make a profit when combined with a simple betting strategy. The authors also used exponential time weighting to discount the effects of past results, while in~\citep{maher1982} the strength of the team is considered to be time-invariant. In~\citep{rue2000}, the authors used a Brownian motion to bind together the strength parameters of the teams in consecutive rounds. The model is then used for betting with a variant of the MPT strategy. \cite{egidi2018combining}, presents a hierarchical Bayesian Poisson model with the scoring rates of the teams being represented by convex combinations of parameters estimated from historical data and betting odds. In \citep{groll2013spain} the authors analyze the explanatory power of bookmakers' odds using pairwise generalized linear mixed Poisson model.}

\rev{
Another modern approach for match outcome predictions are non-parametric and feature-based machine learning models.
\cite{Haghighat2013}, provides a review of machine learning techniques used in outcome predictions of sports events while pointing out some common problems and misconceptions.
In the horse racing domain, a popular logit-based model, combining both ``fundamental features'' and ``odds-derived'' features into a single prediction system, was presented by~\cite{benter2008computer}. This model was also a strong inspiration for the horse racing model evaluated in this paper.
In the domain of soccer, a recent review~\citep{hubacek2019score} discusses a diversity of the common approaches. Notable examples include models from the 2017 Soccer Prediction Challenge~\citep{dubitzky2019}. The winning model from the challenge utilized a boosted tree learner based on an ensemble of score-derived features and simpler ranking and statistical models~\citep{hubacek2019}. This model was also directly used for the soccer betting experiments reported in this paper.
In predictive basketball modelling, it is common to use detailed box-score statistics that are available for the high exposure leagues. Based on diverse features, \cite{Miljkovic2010}, evaluated their model on the NBA, while \cite{Ivankovic2010} used a neural network to predict match outcomes in the League of Serbia. An advanced convolutional neural architecture was then learned over a, so far biggest, set of basketball games in~\citep{hubavcek2019exploiting}. We again directly utilize this basketball model in this paper.
}

\section{Betting Strategies}
\label{sec:strategies}

In the existing literature, the betting strategies range from simple informal techniques, such as flat betting, to the formal approaches, represented mainly by the Markowitz's Modern portfolio theory~\citep{markowitz1952portfolio} and the Kelly criterion~\citep{kelly1956new}, coming from an economical and information-theoretic views of the problem, respectively.

\subsection{Informal Strategies}
\label{sec:strat:informal}

In sports betting practice, most of the focus among punters is being put on the search for outcomes with positive expected value (``value bets''), and the importance of the subsequent investment strategy has often been neglected. Consequently, rather than formal strategies, one can encounter simplistic heuristics such as~\citep{hubacek2017thesis}:

\begin{itemize}
\item Bet a fixed fraction on favourable odds.
\item Bet a fixed fraction on the opportunity with maximal expected value.
\item Bet a fraction equal to the absolute discrepancy between player's and bookmaker's estimates.
\item Bet a fraction equal to the relative discrepancy between player's and bookmaker's estimates.
\item Bet a fraction equal to the estimated probability of winning.
\end{itemize}

% \begin{itemize}
% \item Bet a fixed amount on favorable odds
% \item Bet fixed amount on the opportunity with maximum expected value (max ev).
% \item Bet amount equal to the absolute discrepancy between probabilities predicted by the model and the bookmaker (abs disc bet).
% \item Bet amount equal to the relative discrepancy between probabilities predicted by the model and the bookmaker (rel disc bet).
% \item Bet amount equal to the estimated probability of winning
% \end{itemize}

Lacking any formal foundation, these approaches have been shown generally inferior to the formal strategies, both theoretically and in practice~\citep{hubacek2017thesis}. For completeness, we chose to re-validate the reports by selecting the previously best performing informal strategies of (i) betting fraction w.r.t. the maximal discrepancy (``AbsDisc'') and (ii) betting optimal fraction on the maximal expected value (``MaxEvFrac'') in our experiments (Section~\ref{tab:horses}).

\subsection{Modern Portfolio Theory}
\label{sec:MPT}

Modern Portfolio Theory (MPT) is a standard economic view of the problem based on the idea of the expected value of the profit, possibly transformed by a utility function reflecting the user's particular preferences. The general idea behind MPT is that a portfolio $\bm{f^1}$, i.e. a vector of assets $\bm{f} = f_1, \dots, f_n$, is superior to $\bm{f^2}$, if its corresponding expected profit (Section~\ref{sec:definitions}) is at least as great
\begin{equation}
    \EX[\bm{\rho} \cdot \bm{f^1}] \geq \EX[\bm{\rho} \cdot \bm{f^2}]
\end{equation}
and a given risk measure $risk : \mathbb{R}^n \to \mathbb{R}$ of the portfolio, w.r.t. the given odds, is no greater
\begin{equation}
    risk(\bm{f^1}|\bm{\rho}) \leq risk(\bm{f^2}|\bm{\rho})
\end{equation}
This creates a partial ordering on the set of all possible portfolios. Taking the portfolios that no other portfolio is superior to gives us \rev{a} set of ``efficient portfolios'' $\Theta$~\citep{markowitz1952portfolio}. In simple terms, we trade off the expected $profit-risk$ by maximizing the following
\begin{equation}
    \underset{\bm{f} \in \mathbb{R}^n}{\text{maximize}} ~(\EX[\bm{\rho} \cdot \bm{f}] - \gamma \cdot risk(\bm{f}|\bm{\rho}))
\end{equation}
where $\gamma$ is a hyperparameter reflecting the user's preference for risk.

In the most common setup, the $risk$ of a portfolio $\bm{f}$ is measured through the expected total variance of its profit $Var[\bm{\rho} \cdot \bm{f}] = \bm{f}^T\Sigma \bm{f}$, based on the given covariance matrix $\bm{\Sigma}_n^n$ of net profit of the individual assets. Note that in the case of independent outcomes (Section~\ref{sec:def:outcomes}), this reduces to a diagonal matrix with \rev{the} variance of each binary asset\rev{'s} profit, corresponding to the result $r_i$, following from the given odds $o_i$ and the underlying Bernoulli distribution as
    $\Sigma(i,i) = \hat{P_r}(r_i) \cdot (1-\hat{P_r}(r_i)) \cdot \rho_{i,i}^2$.
MPT can generally thus be expressed as the following maximization problem
\begin{equation}
\label{eq:MPT}
\begin{aligned}
& \underset{\bm{f} \in \mathbb{R}^n}{\text{maximize}}~
& & \EX[\bm{\rho}\cdot\bm{f}]  - \gamma \cdot \bm{f}^T\Sigma \bm{f}\\
& \text{subject to}
& & \sum_{i=1}^{n} f_i = 1, \; f_i \geq 0
\end{aligned}
\end{equation}

Apart from the variance $Var[\bm{w}]$ of the potential net returns $\bm{w} = \bm{\rho} \cdot \bm{f}$, different risk measures have been proposed~\citep{markowitz1952portfolio}, such as standard deviation $\sigma(\bm{w}) = \sqrt{Var[\bm{w}]}$ and coefficient of variation $CV(\bm{w}) = \frac{\sigma(\bm{w})}{\EX[\bm{w}]}$. Generally, there is no \rev{agreed-upon} measure of risk and the choice is thus left to the user.

The MPT approach is often criticized for the disputable choice of risk, which can be perceived as a formal weakness of the approach~\citep{peters2016evaluating}, since in many domains the risk is not easy to define. Moreover, the direct maximization of expected profit can be misleading in games, where the distribution of potential profits is highly skewed, i.e. where the mean profit is very different from the median. This situation naturally occurs in the multiplicative dynamics setting, where maximization of expected value may lead to undesirable outcomes~\citep{peters2016evaluating}.

\subsubsection{Maximum Sharpe Strategy}
\label{sec:MaxSharpe}

Apart from the choice of the risk measure, the inherent degree of freedom in MPT is how to select a particular portfolio from the efficient frontier $\Theta$ (based on the choice of $\gamma$). Perhaps the most popular way to avoid the dilemma is to select a spot in the pareto-front with the highest expected profits w.r.t. the risk. For the risk measure of $\sigma(\bm{w})$, this is known as the ``Sharpe ratio'', generally defined as
\begin{equation}
\frac{\EX[\bm{w}] - r_f}{\sigma(\bm{w})}
\end{equation}
where $\EX[\bm{w}]$ is the expected return of the portfolio, $\sigma(\bm{w})$ is the standard deviation of the return, and $r_f$ is a ``risk-free rate''. Since there is no risk-free investment in sports betting, we can neglect it and reformulate the optimization problem as
\begin{equation}
\begin{aligned}
& \underset{\bm{f} \in \mathbb{R}^n}{\text{maximize}}
& & \frac{\EX[\bm{\rho} \cdot \bm{f}]} {\sqrt{\bm{f}^{T}\bm{\Sigma}\bm{f}}} \\
& \text{subject to}
& & \sum_{i=1}^{n} f_i = 1, f_i \geq 0
\end{aligned}
\end{equation}
the solution of which we will further refer to as the ``MSharpe'' strategy.

The variance-based choices of risk have been often criticized as they penalize excess losses as well as excess returns, which is obviously undesirable. Moreover, the calculation of the MaxSharpe solution is also quite sensitive to errors in the probabilistic estimates, and can often be biased towards extreme solutions, requiring some additional form of control\footnote{E.g. a strategy with no wagers placed would have zero variance resulting into an infinite Sharpe ratio.}. Nevertheless\rev{,} it remains a very popular investment practice, which we include in our experiments.

\subsection{Kelly Criterion}
\label{sec:kelly}

The Kelly criterion\rev{~\citep{kelly1956new, thorp2008kelly}} is based on the idea of expected multiplicative growth in the reinvestment setting (Section~\ref{sec:multiplicative}), so that a portfolio $\bm{f}$ is chosen such that the long-term value of the resulting, continuously reinvested, wealth $W_t$ is maximal (in an infinite horizon of $t$). Note that in this scenario we assume that the same portfolio is going to be presented at each time step. For its multiplicative nature, it is also known as the geometric mean policy, emphasizing the contrast to the arithmetic mean approaches based on the expected value.

The two can, however, be looked at similarly with the use of a logarithmic ``utility function'', transforming the geometric into the arithmetic mean, and the multiplicative into the additive setting, respectively. The problem can then be again expressed by the standard means of maximizing the expected value as

\begin{equation*}
\begin{aligned}
& \underset{\bm{f} \in \mathbb{R}^n}{\text{maximize}}
& & \EX[\log(\bm{O} \cdot \bm{f})]\\
& \text{subject to}
& & \sum_{i=1}^{n} f_i = 1, \; f_i \geq 0
\end{aligned}
\end{equation*}
Note that, in contrast to MPT, there is no explicit term for risk here, as the notion of risk is inherently encompassed in the growth-based view of the wealth progression, i.e. the long-term value of a portfolio that is too risky will be smaller than that of a portfolio with the right risk balance (and similarly for portfolios that are too conservative).

The calculated portfolio is then provably optimal, i.e. it accumulates more wealth than any other portfolio chosen by any other strategy in the limit of $t$. However, this strong result only holds given, considerably unrealistic, assumptions~\citep{kelly1956new, thorp2008kelly, peters2016evaluating}. Similarly to MPT, we assume to know the true probability distribution of game outcomes, and additionally we assume that:
\begin{enumerate}
    \item we are repeatedly presented with the same games.
    \item we play for an infinite amount of time.
\end{enumerate}
Despite the fact that these conditions are impossible to meet in practice, the Kelly strategy is very popular, and its various modifications (Section~\ref{sec:risk}) are prevalent among bettors in practice.

\subsubsection{Quadratic Approximation}
\label{sec:quadraticapprox}

Exact numerical calculation of the Kelly strategy is often \rev{time-consuming}, especially when numerous runs through a large dataset of games is necessary. A practical approach to this issue has been proposed~\citep{busseti2016risk} based on a quadratic approximation of the Kelly's logarithmic utility using the Taylor series expansion. Let us first recall the following.
\begin{equation}
    \log(1+x) = x - \frac{x^{2}}{2} + \dots
\end{equation}
% Now we define a modification of the matrix $\bm{O}$ to be $\bm{\rho}$ as follows.
% \begin{equation}
%     \bm{O} - 1 = \bm{\rho}
% \end{equation}
Next, following~\citep{busseti2016risk}, we make an assumption for the Taylor approximation that our net profits are not too far from zero $\bm{\rho}\cdot{\bm{f}} \approx \bm{0}$ and express the logarithmic part of the Kelly criterion as follows~\citep{busseti2016risk}.
\begin{equation}
    \log(\bm{O} \cdot \bm{f}) = \log(1 + \bm{\rho} \cdot \bm{f})
\end{equation}
allowing us to proceed with the Taylor expansion as
\begin{equation}
    \log(1 + \bm{\rho} \cdot \bm{f}) = \bm{\rho} \cdot \bm{f} - \frac{(\bm{\rho} \cdot \bm{f})^{2}}{2} + ...
\end{equation}
Now taking only the first two terms from the series we transform the expectation of logarithm into a new problem definition as follows
\begin{equation}
\begin{aligned}
& \underset{\bm{f \in \mathbb{R}^n}}{maximize}
& & \EX[\bm{\rho} \cdot \bm{f} - \frac{(\bm{\rho} \cdot \bm{f})^{2}}{2}] \\
& \text{subject to}
& & \sum_{i=1}^{n} f_i = 1.0, ~f_i \geq 0
\end{aligned}
\end{equation}
We will further refer to this strategy as the ``Quadratic Kelly''.
Note that, interestingly, the problem can now be rewritten to
\begin{equation}
\begin{aligned}
& \underset{\bm{f} \in \mathbb{R}^n}{\text{maximize}}
& & \EX[\bm{\rho} \cdot \bm{f}] - \frac{1}{2}\EX[\bm{f}^T (\bm{\rho} \cdot \bm{\rho}^T) \bm{f}] \\
\end{aligned}
\end{equation}
corresponding to the original MPT formulation from Equation~\ref{eq:MPT} for the particular user choice of $\gamma=\frac{1}{2}$.
It follows from the fact that the geometric mean is approximately the arithmetic mean minus $\frac{1}{2}$ of variance~\citep{markowitz1952portfolio}, providing further insight into \rev{the} connection of the two popular strategies of Kelly and Markowitz, respectively.

% \section{Strategy Modifications}
% \label{sec:solutions}

% In this section we review various modifications to the introduced mathematical strategies (Section~\ref{sec:strategies}) that have been proposed to solve different facets of the problem (Section~\ref{sec:definitions}) that emerge in practice. These include computational approximations and various enhanced risk management remedies, stemming from recognizing the uncertainty in the probabilistic estimates.

% \subsection{Multiplicity of Outcomes}
% \todo{hod sem reseni jak se s tim popere Kelly z diplomky?}

% \subsection{Repetition and Dynamics}
% \todo{tady by se hodil treba ten mean asset, mas ho?}

\section{Risk Management Practices}
\label{sec:risk}

The core issue with the mathematical strategies is that their calculations are carried out as if the true probability distribution over the outcomes was known. Moreover\rev{,} they are often sensitive to even \rev{the slightest} error in the estimates. Here we review simple remedies that have been proposed on top of the original strategies to manage the extra risk stemming from the underlying errors, as well as more sophisticated techniques incorporating the uncertainty of estimates directly into computation of \rev{the} strategies.

\subsection{Maximum bet limit}
\label{sec:limit}

Constraining the maximal wager to a fixed value $m$ is probably the most trivial risk-avoiding technique one can encounter, which is probably also why it is the most prevalent one in practice. Moreover, the maximum bet limit often comes from the side of the bookmaker, too, constraining the risk he undertakes w.r.t. each bettor. We thus include this empirical method in our portfolio to see if saturating the invested amount by a fixed threshold might actually improve the overall wealth progression of the existing strategies if properly tuned.

% \todo{nejaky zdroj by to chtelo...}
% \todo{will do}

\subsection{Fractional Approaches}
\label{sec:fractional}

Fractioning is an example of a simple heuristic that is nevertheless very efficient in practice.
The main idea behind any ``fractional approach'' is to bet only a fraction $\omega$ of the calculated portfolio and leave the rest of $1-\omega$ in the cash asset for security. We define such a trade-off index $\omega$ for a portfolio as

\begin{equation}
    \bm{f}_\omega = \omega \bm{f}_{1..n-1} + (1-\omega) \bm{f}_n
\end{equation}
where $\bm{f}_{1..n-1}$ corresponds to the risky part of the portfolio with stochastic assets and $\bm{f}_n$ is the cash asset, as introduced in Section~\ref{sec:def:outcomes}.

The fractional approach is mostly used with the Kelly strategy~\citep{maclean2011growth, thorp2011understanding}, where for $\omega = 0.5$ it is famously referred to as ``half-kelly'' by practitioners. \rev{Nevertheless,} the choice of $\omega$ should depend on the actual distributions and preferences for risk. The same idea of taking only a fraction of the calculated portfolio can generally be applied to any strategy, including MPT, and it is overall useful whenever our estimates are erroneous.

\subsection{Drawdown Constraint}
\label{sec:drawdown}

A drawdown represents a more involved technique that actually modifies the original optimization problem.
The idea of drawdown is to incorporate a special probabilistic constraint into the Kelly strategy so as to push the solution away from the more risky region near the ruin boundary. The choice of the boundary is then left to the user's preference as an input parameter into the optimization problem. The probabilistic boundary is expressed as the following constraint

\begin{equation}
    P(W_t^{min} < \alpha) \leq \beta
\end{equation}
expressing that the probability of our wealth falling below $\alpha$ can be at most $\beta$. 

For the Kelly criterion, following the calculations from~\citep{busseti2016risk}, the constraint is approximately satisfied if the following condition holds
\begin{equation}
    \EX[(\bm{O} \cdot \bm{f})^{-\lambda}] \leq 1 \hspace{5pt} \text{where} \hspace{5pt}  \lambda = \log(\beta) / \log(\alpha)
\end{equation}
Which we can reformat as
\begin{equation}
    \log(\sum_{i=1}^{n} p_i \cdot (\bm{o_i}\cdot f_i)^{-\lambda}) \leq \log(1)
\end{equation}
which can be further simplified~\citep{busseti2016risk} into the following constraint
\begin{equation}
    \log(\sum_{i=1}^{n} \exp(\log(p_i \cdot (\bm{o_i}\cdot f_i)^{-\lambda}))) \leq 0
\end{equation}
which we can finally use in a convex optimization program.

\subsection{Distributionally Robust Optimization}
\label{sec:dro}

Distributionally robust optimization (DRO) can be understood as a stochastic game between a player and nature, where nature picks a distribution $P_r$ from some predefined ambiguity set of distributions $\bm{\Pi}$ so as to inflict maximum damage to the player's utility. This fits quite naturally the setting of sports betting against a fixed-odds bookmaker, where, given the opposing utilities of both, the bookmaker (nature) sets up the odds so as to minimize player's chances for profit.
Generally, DRO is \rev{a} paradigm for decision making under uncertainty where:
\begin{enumerate}
    \item The uncertain problem inputs are governed by a distribution that is itself subject to uncertainty.
    \item The distribution is then assumed to belong to an ambiguity set $\bm{\Pi}$.
    \item The ambiguity set contains all distributions that are compatible with the player's prior information.
\end{enumerate}
Being aware of the uncertainty in her own estimates $P_p = \hat{P_r}$, the player now modifies the optimization problem to account for the worst possible scenario within the given ambiguity set $\Pi$.
\begin{equation*}
\begin{aligned}
& \underset{\bm{f} \in \mathbb{R}^n}{\text{maximize}}
& & \underset{\bm{p} \in \bm{\Pi}}{min} \sum_{i=1}^{n} {p_i} \cdot log(\bm{O_i} \cdot \bm{f})\\
& \text{subject to}
& & \sum_{i=1}^{n} f_i = 1, \; f_i \geq 0
\end{aligned}
\end{equation*}

The ambiguity set $\bm{\Pi}$ can be defined in a number of ways. In~\citep{sun2018distributional}, multiple definitions are explored in connection to the Kelly strategy, such as Polyhedral, Ellipsoidal, or Divergence based. In this review\rev{,} we further narrow our focus to the polyhedral ambiguity set, referred to as the ``box'' uncertainty set, which can be defined as

\begin{equation}
    \bm{\Pi} = \{p_i \hspace{3pt} | \hspace{3pt} |p_i - P_p(r_i)| \leq \eta \cdot P_p(r_i),~\sum_{i=1}^{n} p_i = 1, p_i \geq 0\}
\end{equation}
i.e. constraining each probability $p_i$ to differ by up to a factor of $\eta$ from the nominal player's estimate $P_p(r_i)$ of the probability of result $\mathrm{R}=r_i$.

\section{Experiments}
\label{sec:experiments}

The main purpose of this review is to assess \rev{the} performance of the individual strategies (Section~\ref{sec:strategies}) and their risk modifications (Section~\ref{sec:risk}) in various realistic settings (Section~\ref{sec:definitions}) on real data.
We recall the used strategies, describe the datasets, evaluation protocol, and discuss the conducted experiments with their results.

The strategies for the experiments were chosen with the aim to represent the diverse portfolio of approaches occurring in practice, with the goal to provide an unbiased statistical assessment of their performance limits. The particular strategies chosen with their respective hyper-parameters are specified in Table~\ref{tab:strategies}.

\begin{table}[h!]
\begin{center}
\begin{tabular}{ |c|c|c| }
 \hline
 \textbf{Strategy} & Description & {Hyperparameters}\\
 \hline
 AbsDisc & absolute discrepancy bet (Section~\ref{sec:strat:informal}) & None \\ 
 \hline
 MaxEvFrac & max. EV outcome with fractioning (Section~\ref{sec:strat:informal}) & $\omega \in [0,1]$ \\ 
 \hline
 Kelly & original Kelly strategy (Section~\ref{sec:kelly}) & None \\ 
 \hline
 MSharpe & original max. Sharpe ratio (Section~\ref{sec:MaxSharpe}) & None \\ 
 \hline
 KellyFrac & Kelly strategy with fractioning (Section~\ref{sec:fractional}) & $\omega \in [0,1]$ \\ 
 \hline
 MSharpeFrac & max. Sharpe with fractioning & $\omega \in [0,1]$ \\ 
 \hline
 KellyFracMax & Kelly with fractioning and limiting (Section~\ref{sec:limit}) & $\omega \in [0,1]$, $m \in [0,1]$. \\
 \hline
 MSharpeFracMax & max. Sharpe with fractioning and limiting & $\omega \in [0,1]$, $m \in [0,1]$. \\
 \hline
 KellyDrawdown & Kelly with the drawdown constraint (Section~\ref{sec:drawdown}) & $\alpha$, $\beta \in [0,1]$ \\
 \hline
 KellyRobust & Kelly with distributionally robust optimization & $\eta \in [0,1]$. \\
 \hline
\end{tabular}
\end{center}
\caption{Evaluated strategies and their hyperparameters}
\label{tab:strategies}
\end{table}

%The range of hyperparameters is as follows: $\alpha,\beta,\omega, \eta, m \in [0,1]$

% \begin{itemize}
%     \item AbsDisc - Bet fraction equal to the absolute discrepancy between probabilities predicted by the model and the bookmaker.
%     \item MaxEvFrac - Bet fixed fraction $\omega$ on the opportunity with the maximum expected value.
%     \item Kelly - Choose portfolio according to the Kelly optimization problem in \ref{sec:kelly}.
%     \item MSharpe - Choose portfolio that optimizes Sharpe ratio of selected opportunities, solving the problem specified in \ref{sec:MaxSharpe}
%     \item KellyFrac - Bet fraction $\omega$ of the Kelly portfolio.
%     \item MaxSharpeFrac - Bet fraction $\omega$ of the MSharpe portfolio.
%     \item KellyFracMax - Bet fraction $\omega$ of the Kelly portfolio and limit maximum bet $m$ on a single opportunity.
%     \item MaxSharpeFracMax - Bet fraction of the MSharpe portfolio and limit the maximum bet on a single opportunity.
%     \item KellyDrawdown - Choose Kelly portfolio that satisfies the drawdown constraint. \ref{sec:drawdown}
%     \item KellyRobust - Choose Kelly portfolio solving distributionally robust optimization problem defined in \ref{sec:dro}
% \end{itemize}

\subsection{Datasets}
\label{sec:datasets}

We collected 3 datasets of different properties from 3 different sports - horse racing, basketball, and football, each containing a significant number of ``matches'' \rev{(races and games)} for statistical evaluation. Each of the datasets is further accompanied with realistic models' predictions tuned specifically for each domain. Since our focus here is purely on the betting strategies, we do not elaborate on the models in details beyond their predictive performances, which naturally influence the performance of the strategies, too.
For each of the datasets, we present the following key properties.

\begin{itemize}
    \item $size$ - Dataset size (i.e. \rev{the} number of matches).
    \item $acc_b$ - Accuracy of the bookmaker $b$.
    \item $acc_p$ - Accuracy of the player $p$ (i.e. the predictive model).
    \item $n$ - Number of possible match outcomes ($n=|R|$).
    \item $odds$ - Range of the offered odds.
    \item $margin$ - Average margin present in the odds.
    \item $A_{KL}$ - Kullback-Leibler advantage of the player.
\end{itemize}
The $A_{KL}$ is a statistical measure of \rev{the} difference of the predictive performances (\rev{cross-entropy}) of the player and the bookmaker, respectively. The metric was chosen as it plays a key role in \rev{the} performance of the original Kelly strategy, where the growth of profit can be proved directly proportional to the KL advantage~\citep{cover2012elements}.

\subsubsection{Horse Racing}
\label{sec:horses}

The data for horse racing were collected from the Korean horse racing market (KRA) and provide $2700$ races. The target market of the dataset is the ``win pool'', representing betting for the horse winning the race. The schedule and participation of individual horses in the races varies considerably. Moreover, there is a varying total number of horses, and thus outcomes $n$, in each race, creating \rev{an} interesting challenge for the strategies. We thus assume each race as a completely independent investment opportunity and optimize the strategies accordingly. The model used was a form of conditional logistic regression over various features of the horses \rev{(Section~\ref{sec:related:model})}. The particular dataset properties are specified in Table~\ref{tab:horses}.

\begin{table}[h!]

\begin{center}
\begin{tabular}{ |c|c|c|c|c|c|c|c|}
 \hline
 \textit{size} & \textit{$acc_p$} & \textit{$acc_b$} & $n$ & $odds$ & $margin$ &$A_{KL}$\\ 
 \hline
 $2700$ & $0.512$  & $0.503$ & $\in [6, 16]$ & $\in [1.0, 931.3]$ & $0.2$  & $\approx 0.0022$ \\ 
 \hline
\end{tabular}
\end{center}
\caption{Horse racing dataset properties}
\label{tab:horses}
\end{table}

The specifics of the horse racing dataset come mainly from the fact that it actually originates from a parimutuel market, meaning that the wagers are put into a shared pool from which a certain portion is removed as a profit for the house (margin). Nevertheless\rev{,} we convert it into the discussed fixed-odds setting by assuming the last available state of the money pool to get the possible payoffs/odds~\citep{hausch2008efficiency}. As a result, the ``bookmaker's'' estimate in this case is hence made up entirely from public opinion, and is noticeably less accurate. This provides space for statistical models to gain predictive KL-advantage on the one hand, however, on the other hand, the margin is also considerably higher.

\subsubsection{Basketball}
\label{sec:basket}
 
Next domain we selected is basketball, where we collected box score data from matches in the US National Basketball Association (NBA). The dataset consists of $16000$ games ranging from the year $2000$ to $2015$. The NBA league has a regular schedule of the matches, where each team plays repeatedly with every other team in \rev{so-called} ``rounds''. To emulate the market setting in a realistic fashion, we assume rounds as groups of $10$ scheduled matches to repeatedly appear on the market in parallel (Section~\ref{sec:def:parallel}).
The target market here was the ``money-line'', i.e. betting on the winner of each match. The specifics of the data then comes from the fact that there are only 2 outcomes in the game, directly corresponding to the most basic \rev{coin-tossing} setup of the problem (Section~\ref{sec:definitions}).
The model used was a convolutional neural network based on detailed statistics of the individual players and teams~\citep{hubavcek2019exploiting}. The odds then come from the closing line of the Pinnacle~\footnote{https://www.pinnacle.com/} bookmaker. Notice that in this case the model is not as accurate as the bookmaker, and is thus in a general KL-disadvantage. The particular dataset properties are specified in Table~\ref{tab:basket}.

\begin{table}[h!]
\begin{center}
\begin{tabular}{ |c|c|c|c|c|c|c| } 
 \hline
  \textit{size} &\textit{$acc_p$} & \textit{$acc_b$} & $n$ & $margin$ & $odds$ & $A_{KL}$\\ 
 \hline
 $16000$ & $0.68$  & $0.7$ & $2$ & $0.038$ & $\in [1.01, 41]$  & $\approx -0.0146$\\
 \hline
\end{tabular}
\end{center}
\caption{Basketball dataset properties}
\label{tab:basket}
\end{table}

\subsubsection{Football}
\label{sec:football}

The football dataset consists of $32000$ matches collected from various leagues all over the world. The schedule in each football league is similar in spirit to that of \rev{the} NBA, and so we again assume the market setting with $10$ parallel games (Section~\ref{sec:def:parallel}). The target market was again money-line betting. The outcomes in football include a draw, resulting \rev{in} a moderate $n=3$ setting. Interestingly, the original dataset~\citep{dubitzky2019} contained merely the historical results of the matches, and the model has thus been built purely from score-derived features. Particularly, the model was a form of gradient-boosted trees learner, winning the 2017's Soccer prediction challenge~\citep{dubitzky2019}. The odds were again provided by \rev{Pinnacle but, this time, we} took the more \rev{favourable} opening line. Despite varying over different leagues, the overall margin is slightly lower than in basketball, and the model in a slightly lower, yet still considerable, KL disadvantage. The particular dataset properties are specified in Table~\ref{tab:football}.

\begin{table}[h!]
\begin{center}
\begin{tabular}{ |c|c|c|c|c|c|c| } 
 \hline
  \textit{size} &\textit{$acc_p$} & \textit{$acc_b$} & $n$ & $margin$ & $odds$ & $A_{KL}$\\ 
 \hline
 $32000$ & $0.523$  & $0.537$ & $3$ & $0.03$ & $\in [1.03, 66]$  & $\approx -0.013$\\ 
 \hline
\end{tabular}
\end{center}
\caption{Football dataset properties}
\label{tab:football}
\end{table}

\subsection{Evaluation Protocol}
\label{sec:ex:protocol}

The models providing the probabilistic estimates were trained following the natural order of the matches in time, so that all of their estimates are actual future predictions, i.e. out-of-sample test outputs for matches unseen in the training phase.

For the actual optimization problems of the individual strategies, we have chosen to work with the cvxpy \citep{cvxpy} as the main optimization framework. For each strategy, we first solved the given problem using the Embedded Conic Solver (ECOS)~\citep{domahidi2013ecos}, and should a numerical problem arise\rev{,} we proceed with solving the problem using the Splitting Conic Solver (SCS)~\citep{o2016scs}.

While many of the chosen strategies (Table~\ref{tab:strategies}) contain hyperparameters to be set, we additionally tuned each for the best possible performance via grid-search, too. The individual hyperparameter ranges for the grid-search can be found in Table~\ref{tab:strategies}.
To provide an unbiased \rev{estimate} of their actual performance in practice, we also followed a strict evaluation protocol for each of the strategies. This means that we have (i) split each dataset into training and testing subsets, (ii) found the best hyperparameter setting on the training subset, and (iii) evaluated the fixed setting on the test subset.

To make the output profit measures (Section~\ref{sec:metrics}) more robust, both the training and testing is evaluated by generating $1000$ separate ``runs'' through each subset, where the sequence of games is randomly reshuffled and $10\%$ of games are randomly removed each time (the split between train and test always remains respected). We hence evaluate properties of each strategy on $1000$ separate wealth investment trajectories through previously unseen games.

\subsubsection{Hyperparameter Selection}
\label{sec:hyperpar}

To choose the best possible strategy setting on the train set, we looked for hyperparameters with the following criteria 
\begin{equation*}
\begin{aligned}
& {\text{maximize}}
& & median(\bm{W_{f}}) \\
& \text{subject to}
& & Q_{5} > 0.9
% & &  & count(\bm{W_{ruin}}) = 0
\end{aligned}
\end{equation*}
i.e. we always chose a strategy that reached the maximum median final wealth, given that no more than $5\%$ of the wealth trajectories did not fall below $90\%$ of \rev{the} final wealth. Hyperparameter settings that did not meet the required criterion were simply removed from consideration. While the presented hyperparameter selection criteria might seem somewhat arbitrary and could be argued, our aim was to follow the natural desiderata of wealth progression for bettors in practice. That is to mainly prevent the occurrence of ruin (``survival first''), and then maximize the potential profits for the typical (median) bettor.

\subsubsection{Evaluation Metrics}
\label{sec:metrics}

For the actual final evaluation of the strategies on the test set, we chose a range of diverse metrics to provide more insights into the properties of the individual strategies and game settings. The metrics are as follows
\begin{itemize}
    \item $median(W_f)$ - median final wealth position.
    \item $mean(W_f)$ - mean final wealth position.
    \item $min(W_i)$ - lowest wealth position.
    \item $max(W_i)$ - maximal wealth position.
    \item $sigma(W_f)$ - standard deviation of \rev{the} final wealth positions.
    \item $ruin$ \% - ruin percentage of wealth trajectories
\end{itemize}
for which we define a $ruin$ situation as falling below $0.01\%$ of the initial bank $W_0$ at least once during the entire investment period. Note that as opposed to the original definition of ruin in the Kelly strategy~\citep{kelly1956new}, we have chosen a small \textit{non-zero} threshold, since in practice there is a low amount of money effectively causing \rev{the} inability to place a minimal bet, which is a constraint often present in the market.

\subsection{Results}
\label{sec:results}

\begin{figure}[h!]
\includegraphics[width=0.85\textwidth]{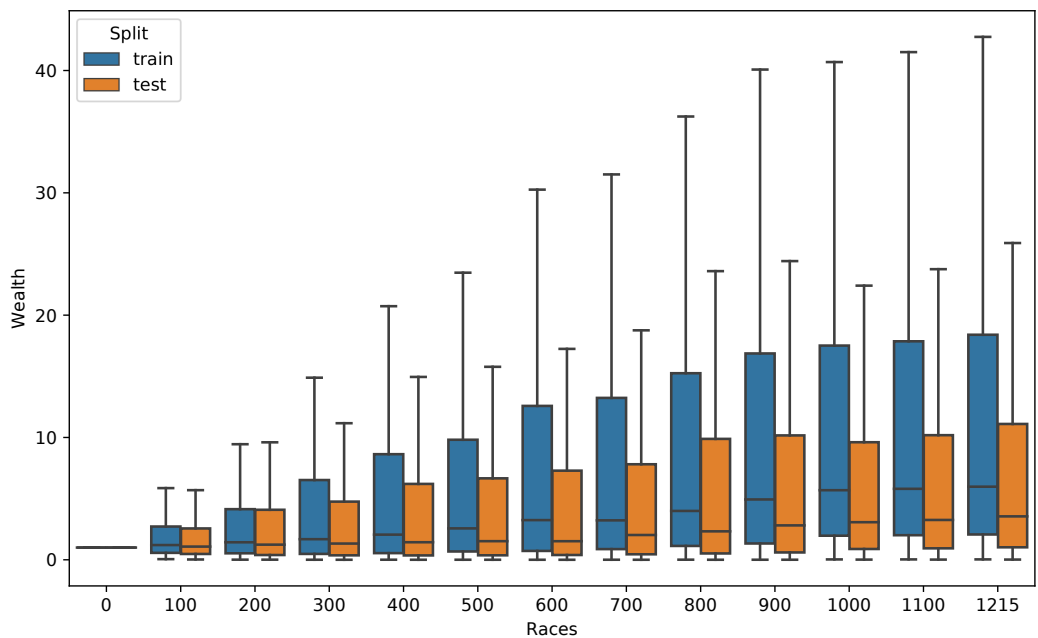}
\centering
\caption{Wealth progression of the KellyFracMax strategy in the horse racing scenario (Section~\ref{sec:horses}).}
\label{fig:horses}
\end{figure}

\begin{figure}[h!]
\includegraphics[width=0.85\textwidth]{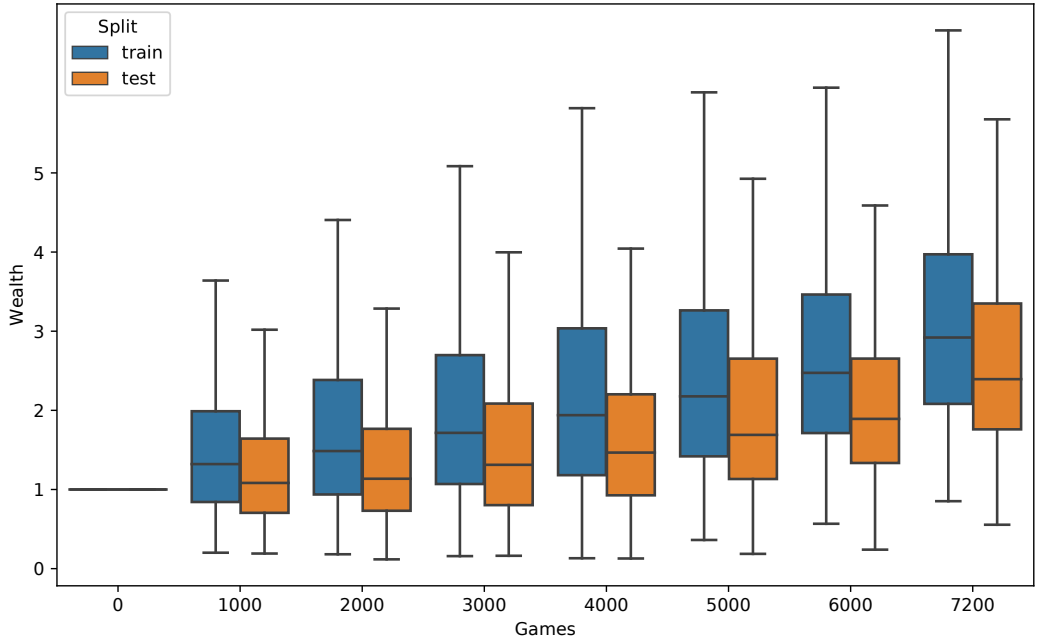}
\centering
\caption{Wealth progression of the KellyFrac strategy in the basketball scenario (Section~\ref{sec:basket}).}
\label{fig:basket}
\end{figure}

\begin{figure}[h!]
\includegraphics[width=0.85\textwidth]{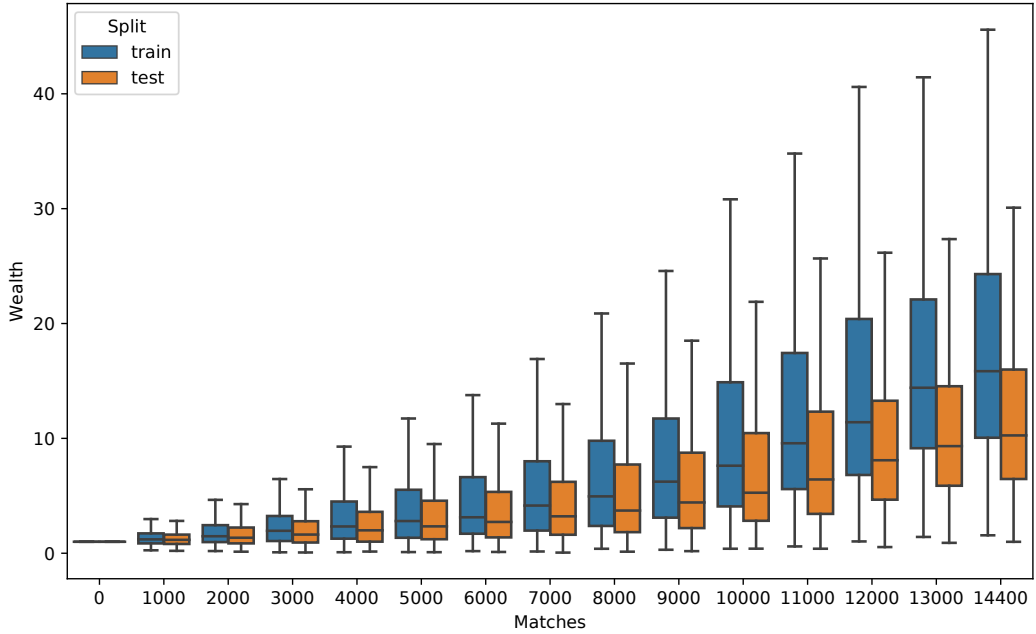}
\centering
\caption{Wealth progression of the KellyDrawdown strategy in the football scenario (Section~\ref{sec:football}).}
\label{fig:football}
\end{figure}

Finally\rev{,} we present performances (Section~\ref{sec:metrics}) of the individual strategies (Section~\ref{sec:experiments}) over each of the datasets (Section~\ref{sec:datasets}). Apart from the evaluation metrics in the final state of wealth progression $W_{f}$, we present the summarized wealth progression trajectories for a selected ``best'' strategy with maximal median final wealth for each of the datasets, to demonstrate the overall bankroll dynamics. \rev{The evaluation metrics for horse racing, basketball, and football datasets are presented in Table~\ref{experiments:metrics:horses}, Table~\ref{experiments:metrics:basketball}, and Table~\ref{experiments:metrics:football}, respectively. The wealth progression trajectories for the best strategies are then displayed in
Figure~\ref{fig:horses}, Figure~\ref{fig:basket} and Figure~\ref{fig:football}, respectively.}
\begin{table}[h!]
\begin{center}
\begin{tabular}{ |c|c|c|c|c|c|c|c|}
 \hline
 \textit{\textbf{strategy}} & $median(W_f)$ & $mean(W_f)$ & $min(W_i)$ & $max(W_i)$ & $sigma(W_f)$ & $ruin$ \%\\
 \hline
 AbsDisc & 0.0019 & 0.03 & 4e-08 & 27.1 & 0.04 & 85.2 \\
 \hline
 MaxEvFrac & 0.86 & 2.13 & 2e-09 & 711 & 4.7 & 36.1 \\
 \hline
 \hline
 Kelly & 4.11 & 15.6 &  7e-05 & 2167.8 & 59.8 & 0.6 \\
 \hline 
 MSharpe & 3.92 & 17.8 & 9e-06 & 2231.1 & 48.3 & 12.1 \\
 \hline 
 KellyFrac & 3.39 & 14.2 & 0.003 & 213.2 & 32.1 & 0 \\
 \hline 
 MSharpeFrac & 3.28 & 16.9 & 8e-05 & 253.3 & 26.5 & 0.2 \\
  \hline 
 KellyFracMax & 3.49 & 13.8 & 0.0057 & 168.1 & 29.3 & 0 \\
 \hline 
 MSharpeFracMax & 3.41 & 15.2 & 0.0065 & 194.3 & 25.4 & 0 \\
 \hline 
 KellyDrawdown & 3.3 & 13.7 & 0.009 & 112.4 & 22.4 & 0 \\
 \hline 
 KellyRobust & 2.97 & 4.1 & 0.08 & 77.3 & 7.2 & 0 \\
 \hline
\end{tabular}
\end{center}
\caption{Final wealth statistics of the strategies in the horse racing scenario (Section~\ref{sec:horses}).}
\label{experiments:metrics:horses}
\end{table}

\begin{table}[h!]
\begin{center}
\begin{tabular}{ |c|c|c|c|c|c|c|c|}
 \hline
 \textit{\textbf{strategy}} & $median(W_f)$ & $mean(W_f)$ & $min(W_i)$ & $max(W_i)$ & $sigma(W_f)$ & $ruin$ \%\\ 
 \hline
 Kelly & 9.1e-6 & 1.8e-05 & 1.9e-20 & 3312.2 & 1.7e-05 & 100 \\
 \hline 
 MSharpe & 1.3e-06 & 5.1e-05 & 4.1e-21 & 2911 & 9.7e-06 & 100 \\
 \hline 
 KellyFrac & 2.4 & 2.7 & 0.11 & 24.1 & 1.34 & 0 \\
 \hline 
 MSharpeFrac & 1.24 & 1.97 & 0.002 & 19.6 & 0.85 & 0 \\
  \hline 
 KellyFracMax & 2.3 & 2.5 & 0.13 & 20.9 & 1.27 & 0 \\
 \hline 
 MSharpeFracMax & 1.2 & 1.7 & 0.008 & 12.1 & 0.56 & 0 \\
 \hline 
 KellyDrawdown & 2.21 & 2.9 & 0.14 & 29.1 & 1.3 & 0 \\
 \hline 
 KellyRobust & 1.39 & 1.46 & 0.23 & 10.9 & 0.45 & 0 \\
 \hline
\end{tabular}
\end{center}
\caption{Final wealth statistics of the strategies in the basketball scenario (Section~\ref{sec:basket}).}
\label{experiments:metrics:basketball}
\end{table}

\begin{table}[h!]
\begin{center}
\begin{tabular}{ |c|c|c|c|c|c|c|c|}
 \hline
 \textit{\textbf{strategy}} & $median(W_f)$ & $mean(W_f)$ & $min(W_i)$ & $max(W_i)$ & $sigma(W_f)$ & $ruin$ \%\\ 
 \hline
 Kelly & 2.3e-09 & 5.2e-08 & 1.6e-21 & 5844.2 & 2.7e-07 & 100 \\
 \hline 
 MSharpe & 1.8e-10 & 3.0e-07 & 5.9e-27 & 2617 & 4.2e-07 & 100 \\
 \hline 
 KellyFrac & 10.05 & 11.8 & 0.03 & 182 & 9.7 & 0 \\
 \hline 
 MSharpeFrac & 9.9 & 13.6 & 0.016 & 211 & 9.1 & 0 \\
  \hline 
 KellyFracMax & 10.03 & 11.2 & 0.007 & 144 & 9.2 & 0 \\
 \hline 
 MSharpeFracMax & 10.1 & 13.1 & 0.005 & 193 & 8.7 & 0 \\
 \hline 
 KellyDrawdown & 10.25 & 12.4 & 0.09 & 122 & 9.3 & 0 \\
 \hline 
 KellyRobust & 6.2 & 7.3 & 0.28 & 27.7 & 5.6 & 0 \\
 \hline
\end{tabular}
\end{center}
\caption{Final wealth statistics of the strategies in the football scenario (Section~\ref{sec:football}).}
\label{experiments:metrics:football}
\end{table}

% \textbf{insights}
% \begin{enumerate}
%     \item Bad strategies, specifically the informal ones( MaxEV, Absolute Discrepancy) can ruin you even in advantageous scenario.
%     \item Smart strategy can profit even in a seemingly unprofitable scenario. By typical measures: KL Advantage, count accuracy etc. 
%     \todo{The type of model should be mentioned, obviously not all disadvantegous scenarios can be profitable with good strategy}
%     \item When looking at final median wealth, $median(W_f)$ Kelly based strategies generally work better.
%     \item When considering final mean wealth, $mean(W_f)$  MSharpeFrac, MSharpeFracMax work better. 
%     \item Kelly DrawDown (KellyDrawdown) performs comparably to Kelly Fractional in all tested scenarios.
%     \item Kelly Distributionally Robust (KellyRobust) staked very conservatively compared to all strategies. Very risk averse but stable. Makes sense as it optimizes worst possible case. 
    
% \end{enumerate}

% \vspace{cm}

Firstly, the results of our experiments confirm that the, regularly used, informal betting strategies (Section~\ref{sec:strat:informal}) are clearly inferior to all the formal strategies, in agreement with the previous reports~\citep{hubavcek2019exploiting}. Moreover, they often lead to ruin even in \rev{a} situation with statistical model advantage, as reported for the horse racing dataset in Table~\ref{tab:horses}, for which we decided not to include them further.

As expected, the formal strategies based on Modern Portfolio Theory (MPT) (Section~\ref{eq:MPT}) and Kelly Criterion (Section~\ref{sec:kelly}) performed reasonably in the setting with \rev{a} statistical advantage $A_{KL}$ of having a more precise model. However, since they are based on unrealistic mathematical assumptions, their actual risk profile might be unexpected in practice. Using any of the proposed practices for additional risk management (Section~\ref{sec:risk}) generally led to a considerably lower volatility while keeping the wealth progression of a typical (both mean and median) bettor reasonably high. Also, following the mathematical properties of the pure form of both the strategies, they both lead to a certain ruin in scenarios without statistical $A_{KL}$ advantage of the model, which is exhibited in practice, too (Table~\ref{tab:basket}, Table~\ref{tab:football}).

On the other hand, a smart strategy modification can generate profits even in the statistically disadvantageous scenarios, as measured by the $A_{KL}$. Naturally, this does not hold universally and particular properties of the underlying models must be considered, too, since there are surely disadvantageous scenarios where no strategy can make profits by any means (Example~\ref{ex:coin1}).

The insights from the experiments regarding the discord between the approaches of MPT and Kelly roughly follow the intuitions behind the individual strategies. That is that the strategies based on the Kelly criterion (Section~\ref{sec:kelly}) result in a generally higher \textit{median} final wealth, while strategies based on the MPT (Section~\ref{sec:MPT}) result in a generally higher \textit{mean} final wealth, corresponding to the underlying expected value-based motivation. Interestingly, in the football dataset (Table~\ref{tab:football}) the mean final wealth performance of MPT is slightly lower than that of the Kelly-based strategies. However, we should note that the hyperparameter selection criteria (Section~\ref{sec:hyperpar}) can also be considered slightly biased in \rev{favour} of the Kelly approaches.

From a practical perspective, the drawdown modification of the Kelly criterion (Section~\ref{sec:drawdown}) seemed to perform very similarly to the, much less sophisticated, fractional approach (Section~\ref{sec:fractional}), further supporting its popular use in practice. While the distributionally robust modification of Kelly (Section~\ref{sec:dro}) achieved generally lowest final wealth scores, it was also the overall most stable strategy with the highest minimal final wealth. This is in complete accordance with its pessimistic underlying setting optimizing for the worst case scenario, which might be appealing to highly risk-averse bettors.

\section{Conclusions}
\label{sec:conclusion}

In this experimental review, we investigated the two most prominent streams of betting investment strategies based on the views of the Modern Portfolio Theory and the Kelly criterion, together with a number of their popular modifications aimed at additional risk management in practice, where their original underlying mathematical assumptions do not hold. We tested the strategies on 3 large datasets from 3 different sport\rev{s} domains of horse racing, basketball, and football, following a strictly unified evaluation protocol to provide unbiased estimates of \rev{the} performance of each method while tuning their \rev{hyperparameters}.

The results of our experiments suggest \rev{the} superiority of the formal mathematical approaches over the informal heuristics, which are often used in practice, however\rev{,} the experiments also revealed their weaknesses stemming from the unrealistic mathematical assumptions, particularly the knowledge of the true probability distribution over the \rev{match} outcomes.
\rev{
Consequently, when used in their plain original form, the formal strategies, i.e. the maximum Sharpe and Kelly, proved infeasible in almost all practical scenarios with uncertain probability estimates. Particularly, the theoretically optimal strategies often led to ruin instead of maximal profit, calling for the need of the additional risk management practices.
% The assumptions for optimality of the two formal approaches are almost always broken and the application of one method or the other almost always requires adjustments to the resulting fractions. 
}
\rev{The results of the subsequent modifications of the optimal strategies then suggested that reasonable trade-offs in wealth progression can be found in actual betting practice with the appropriate techniques, even in scenarios with worse model predictions than that of the bookmaker.}

\rev{Based on the experiments, we conclude that, for common practical purposes, the most suitable option out of the strategies reviewed seems to be the fractional Kelly, given that the fraction hyperparameter has been properly tuned to reflect the amount of uncertainty in each particular problem setting. The approach achieved the best, or close to the best, performance as evaluated by the chosen metrics in most of our experiments while being comparatively simpler than the other strategies. Our findings thus further support its common use in betting practice. The other common practice of setting a maximum bet limit was inconclusive as it improved the overall results in some domains (Table~\ref{experiments:metrics:horses}) while decreasing the profits in others (Table~\ref{experiments:metrics:basketball}).}
\rev{The distributionally robust Kelly strategy then proved to be the safest in all of the experiments, and can thus be suggested to extremely risk-averse practitioners. The second safest strategy was then to incorporate the drawdown constraint, which also proved quite efficient in trading of the security for profit.}

\section*{Acknowledgments}

The authors acknowledge support by Czech Science Foundation grant. no 20-29260S.

%%%%%%%%%%%Section B
%%%%%%%%%%%%%%%%%%%

%%%%%%%%%%%%%%%%%bibliography style
% \bibliography{references.bib}
% \bibliographystyle{agsm}
% \bibliographystyle{plain}
% %\bibliographystyle{alpha}
% %\bibliographystyle{unsrt}
% %\bibliographystyle{abbrv}
% \bibliography{references}
%%%%%%%%%%%%%%%%%%%%%%%%
%%%%%%%%%%%%%%%%%%%%%%%%%%%%%%%%%%%

\end{document}